\documentclass[12pt]{iopart}

\usepackage{graphicx}
\begin{document}

\title[Epitaxial Co$_2$Cr$_{0.6}$Fe$_{0.4}$Al thin films and magnetic tunneling junctions]{Epitaxial Co$_2$Cr$_{0.6}$Fe$_{0.4}$Al thin films and magnetic tunneling junctions}

\author{Andr\'es Conca, Martin Jourdan and Hermann Adrian}

\address{Institute of Physics, Johannes Gutenberg University,
Staudinger Weg 7, 55128 Mainz, Germany}
\ead{jourdan@uni-mainz.de}

\begin{abstract}
Epitaxial thin films of the theoretically predicted half metal Co$_2$Cr$_{0.6}$Fe$_{0.4}$Al were deposited by dc magnetron sputtering on different substrates and buffer layers. The samples were characterized by x-ray and electron beam diffraction (RHEED) demonstrating the B2 order of the Heusler compound with only a small fraction of disorder on the Co sites. Magnetic tunneling junctions with Co$_2$Cr$_{0.6}$Fe$_{0.4}$Al electrode, AlO$_x$ barrier and Co counter electrode were prepared. From the Julli\`ere model a spin polarisation of Co$_2$Cr$_{0.6}$Fe$_{0.4}$Al of 54\% at T=4K is deduced. The relation between the annealing temperature of the Heusler electrodes and the magnitude of the tunneling magnetoresistance effect was investigated and the results are discussed in the framework of morphology and surface order based of in situ STM and RHEED investigations.
\end{abstract}

\pacs{75.70.-i, 73.40.Cg, 61.10.Nz}
\maketitle

\section{\label{sec:intro}Introduction}

\begin{figure}
\center
\includegraphics[width=300pt, angle=-90]{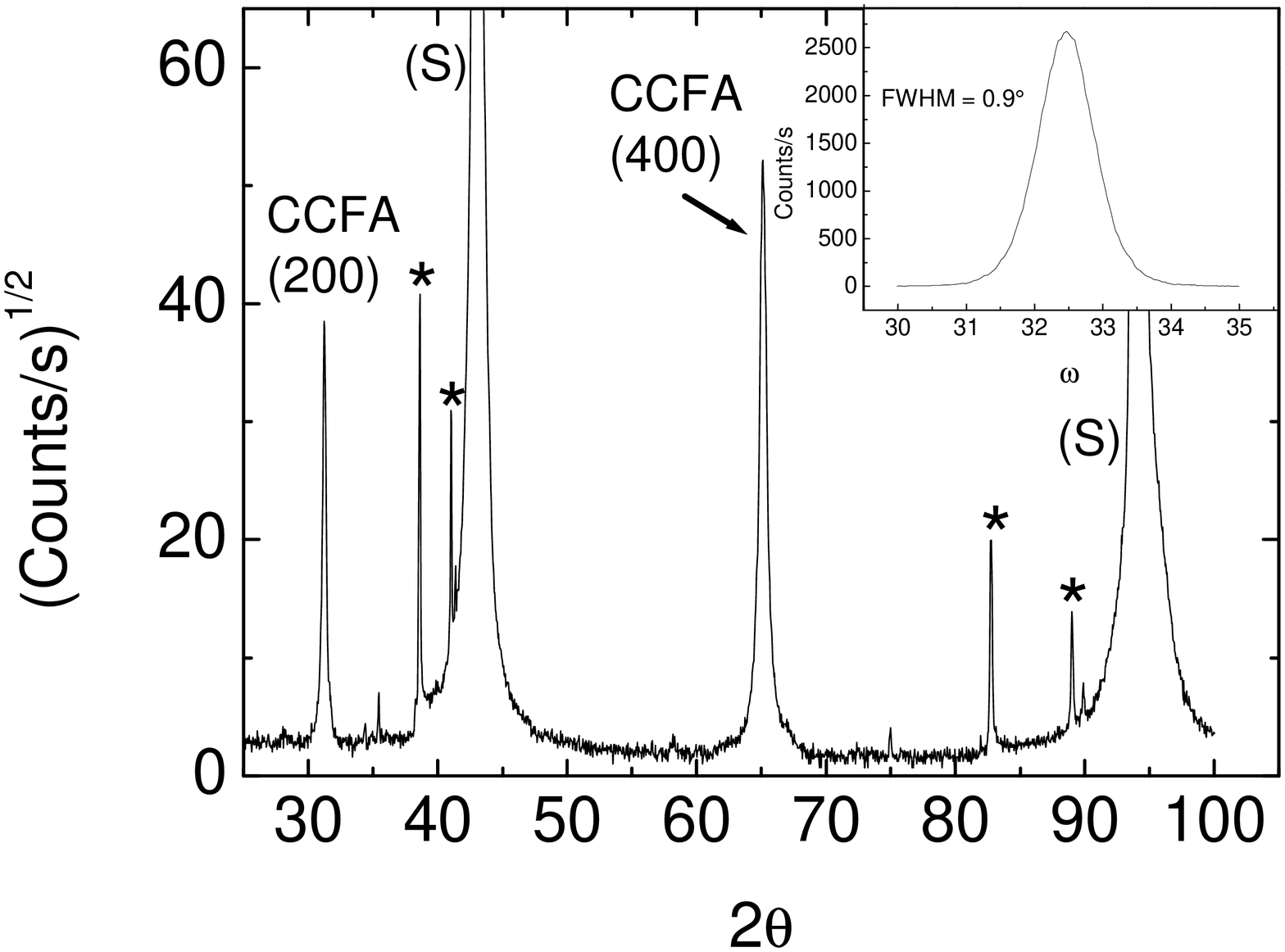}
\caption{\label{mgo} X-ray diffraction $\Theta$/2$\Theta$ scan of a 100 nm CCFA film deposited at 100$^{\rm o}$C on a Fe buffer layer on a MgO(100) substrate and annealed at 600$^{\rm o}$C. (S) are substrate reflections, * are substrate reflections due to secondary x-rays wavelengths. Please note that the square root of the scattering intensity is plotted on the y-axis. The inset shows the rocking curve of the (400) reflection with a full width at half maximum (FWHM) of 0.9$^{\rm o}$.}
\end{figure}

The ferromagnetic full Heusler alloy Co$_2$Cr$_{0.6}$Fe$_{0.4}$Al (CCFA) has attracted great interest in the field of spintronics. The high T$_C$  ($\sim$800K) and the expected half metallicity \cite{fecher, wurmehl} make CCFA  a good candidate for applications in spintronic devices.
Magnetic tunneling junctions (MTJ's) are one example for such a device, which is also useful for the investigation of the electronic states and spin polarisation of the ferromagnetic electrodes. From  the simplifying Julli\`ere model \cite{julliere}, a high tunelling magnetoresistance (TMR) is expected as a result of a high spin polarisation at the Fermi level.

The highest TMR ratios of junctions with CCFA electrode have been reported in conjunction with MgO barriers \cite{marukame}. However, due to the  well known spin filtering effects of ordered MgO barriers \cite{mavropoulos,butler}, this high TMR may be not related to an intrinsic high spin polarisation of CCFA. In order to get information about the total spin polarisation in CCFA, the use of amorphous barriers is advantageous. In this kind of barrier, incoherent scattering at the interfaces samples all the bands and the simple density of states arguments of the Julli\`ere model may be applied. Reports of the use of CCFA in MTJ's with alumina barriers show a clearly reduced TMR ratio in comparison with junctions with MgO barrier. Applying the Julli\`ere model, from junctions with AlO$_x$ barrier a maximum spin polarisation of $\simeq$52\% for Co$_2$Cr$_{0.6}$Fe$_{0.4}$Al at 5K has been deduced \cite{inomata}.

For the spin polarisation of Heusler compounds the degree of structural disorder plays a crucial role. The preference of CCFA to grow in the B2 structure instead of the fully ordered L2$_1$ structure is well known. The B2 structure implies disorder only on the Cr/Fe and Al positions. Calculations predict that this type of disorder has only a minor effect on the spin polarisation. In contrast, disorder on the Co sites is predicted to reduce strongly the spin polarisation \cite{miura}. 

\begin{figure}
\center
\includegraphics[width=300pt, angle=-90]{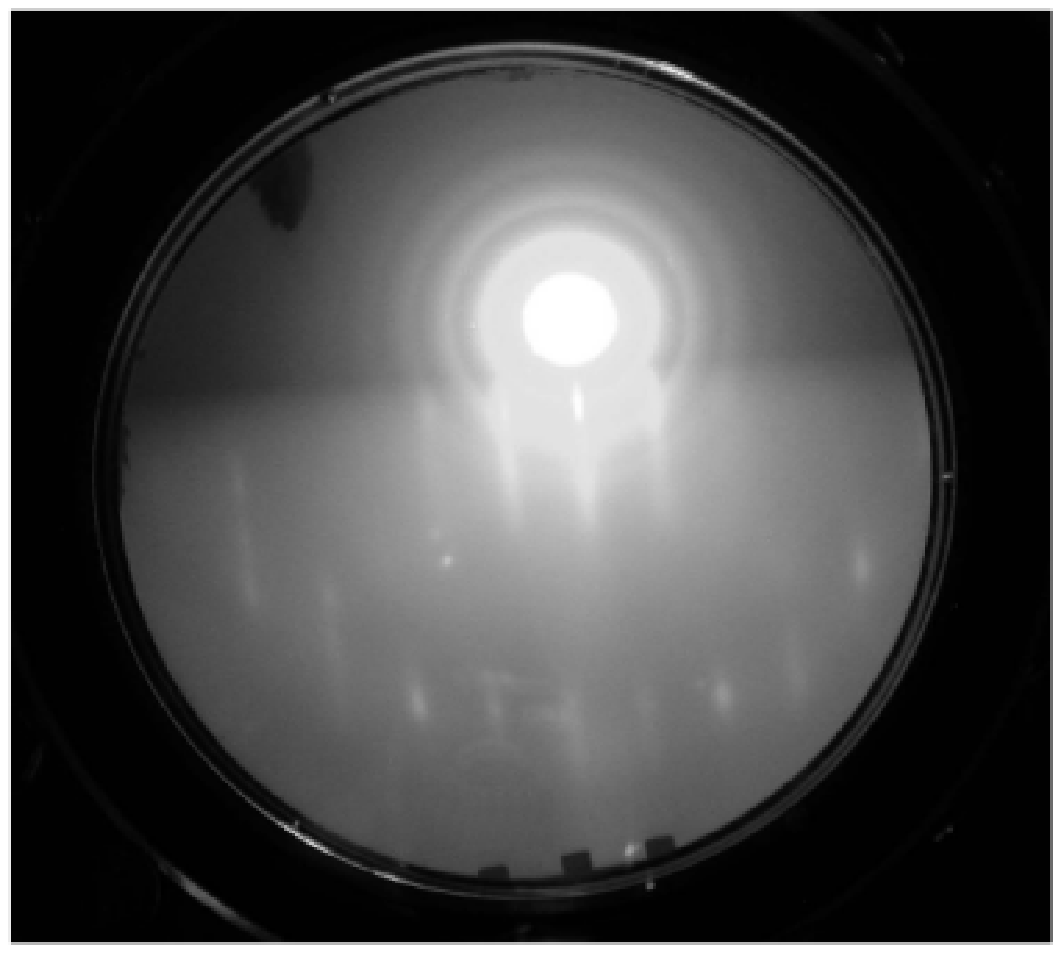}
\caption{\label{rheed-mgo} RHEED pattern of a CCFA thin film deposited on a Fe buffer layer on MgO (100) and after annealing at 600$^{\rm o}$C. The zeroth and first order surface reflections are clearly visible.} 
\end{figure}

\section{\label{sec:prep}Co$_2$Cr$_{0.6}$Fe$_{0.4}$Al thin film preparation}

Epitaxial CCFA films were deposited on Fe or Cr buffer layers on MgO (100) substrates and  directly on Al$_2$O$_3$ (11$\bar{2}$0) and MgAl$_2$O$_4$ (100) substrates. The MgO substrates were annealed ex situ in an oxygen atmosphere at 950$^{\rm o}$C for 2 hours. Additionally, the substrates were cleaned prior to loading in the deposition system by exposing them to a microwave oxygen plasma.  The buffer layers were deposited by an electron beam evaporator in a MBE chamber with a base pressure of $\sim$10$^{-10}$ mbar. The CCFA and Co electrodes as well as the Al for the barrier were deposited by dc magnetron sputtering in a different chamber with a base pressure of $\sim$10$^{-8}$ mbar. Both chambers are connected by a vacuum transfer system.
The CCFA thin films were deposited at a substrate temperature of 100$^{\rm o}$C with an argon pressure of 6.0$\cdot10^{-3}$mbar. The deposition rate was 0.5nm/s. After deposition the CCFA films were annealed for 5 min at 550-600$^{\rm o}$C. 
 
\begin{figure}
\center
\includegraphics[width=300pt, angle=-90]{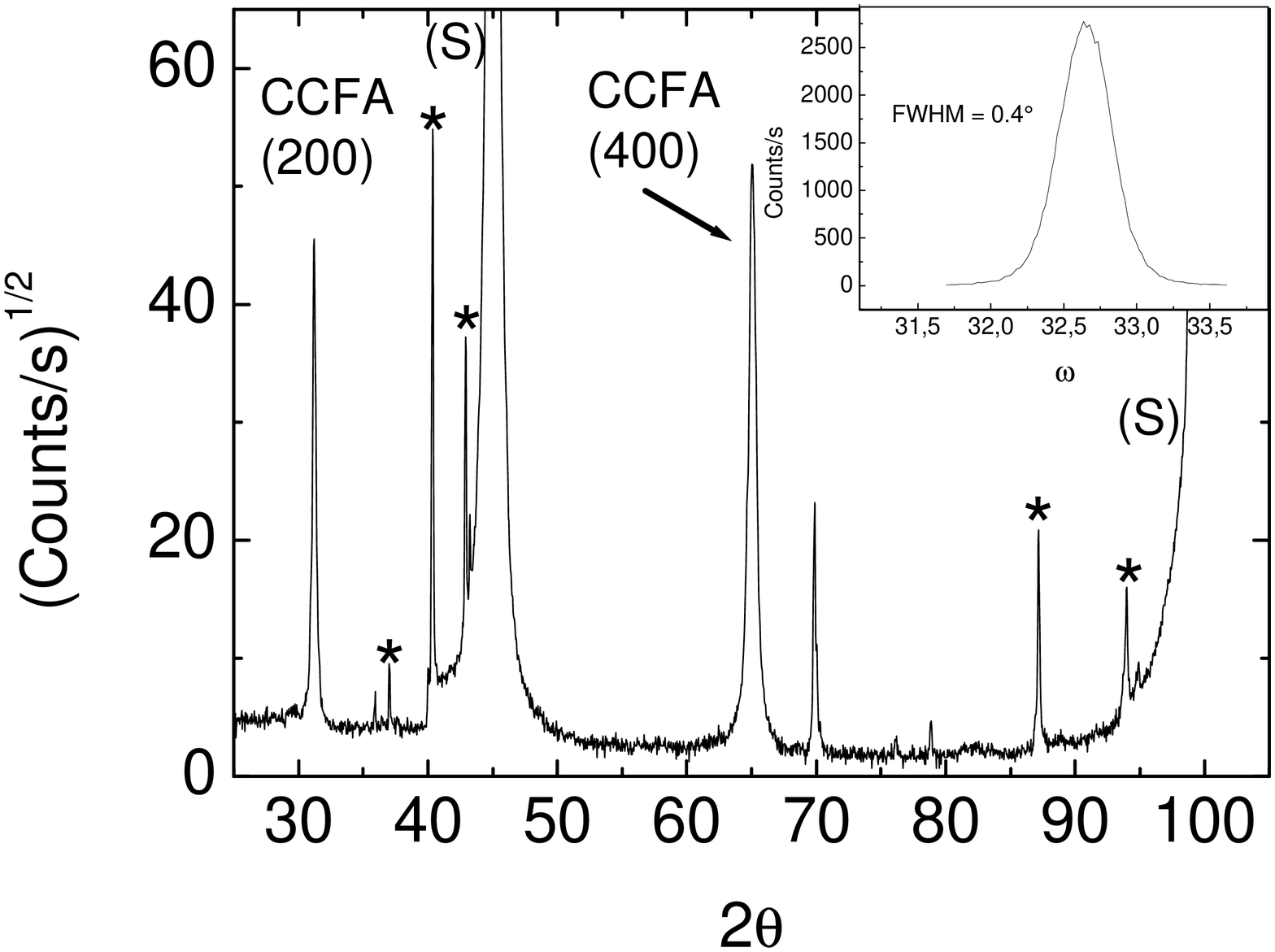}
\caption{\label{mgalo} X-ray diffraction $\Theta$/2$\Theta$ scan of a 100 nm CCFA thin film deposited at 100$^{\rm o}$C on a MgAl$_2$O$_4$ (100) substrate and annealed at 600$^{\rm o}$C. The inset shows the rocking curve of the (400) reflection.}
\end{figure}

\begin{figure}
\center
\includegraphics[width=300pt, angle=-90]{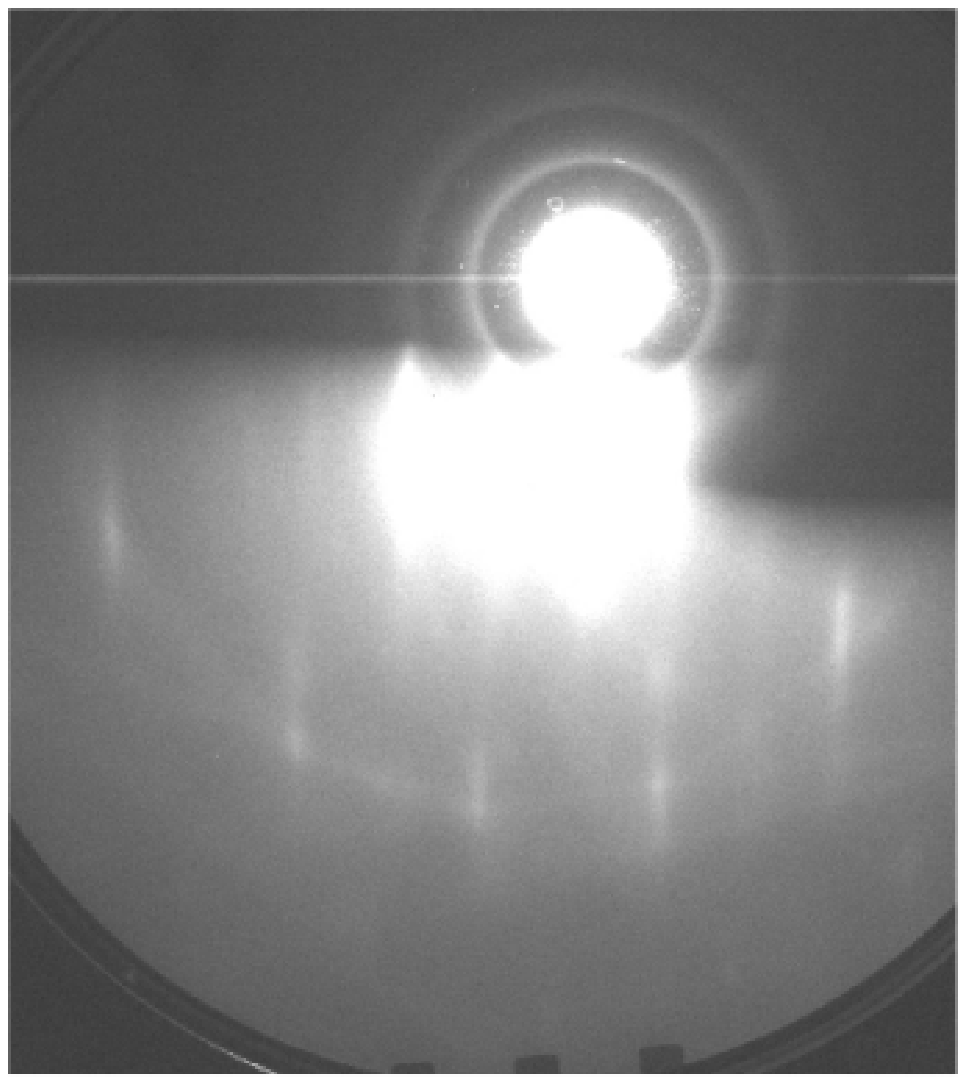}
\caption{\label{rheed-mgalo} RHEED pattern of a CCFA thin film deposited on  MgAl$_2$O$_4$ (100) and annealed at 600$^{\rm o}$C.}
\end{figure}

\begin{figure}
\center
\includegraphics[width=300pt, angle=-90]{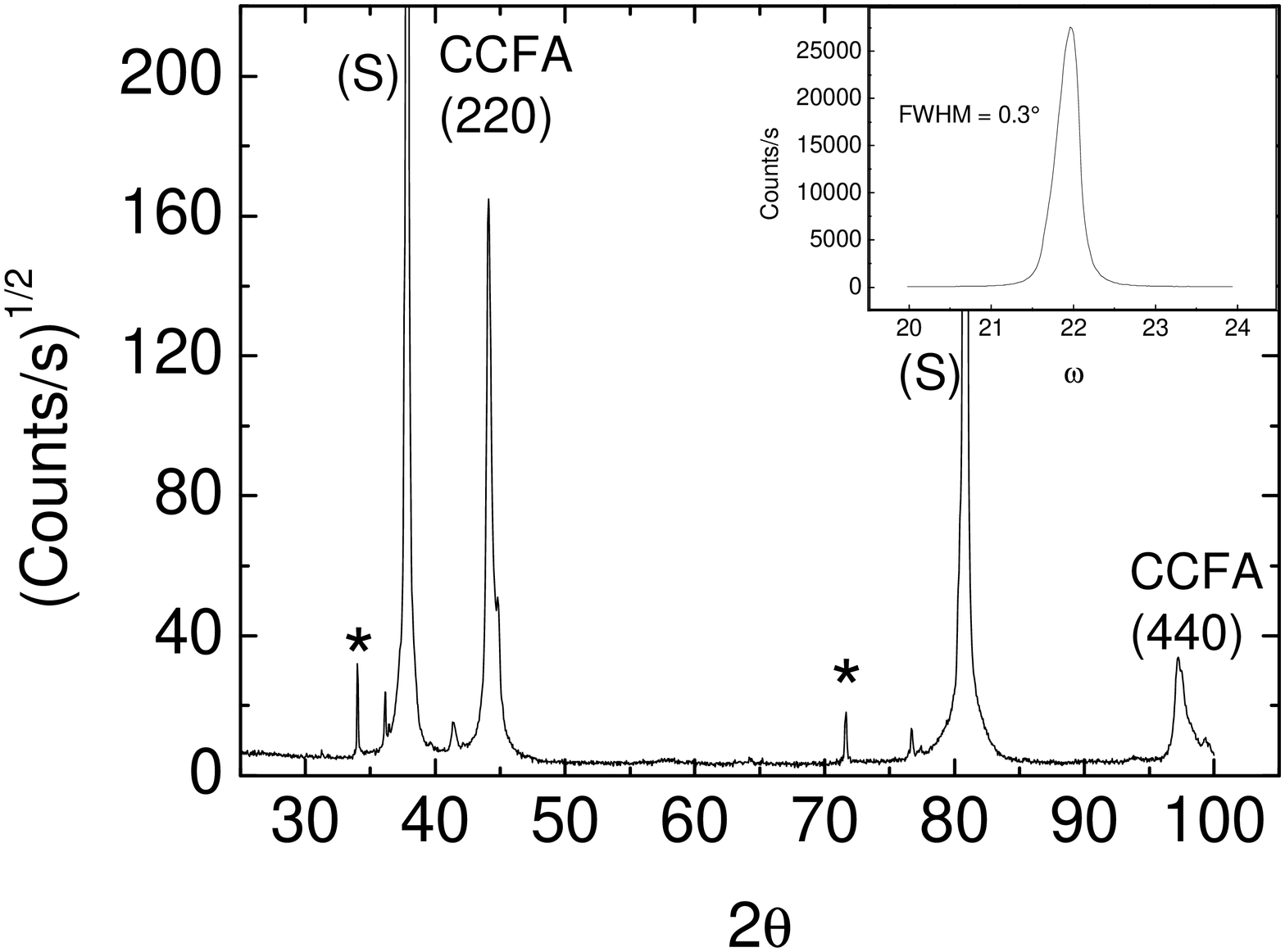}
\caption{\label{alo} X-ray $\Theta$/2$\Theta$ scan of a 100 nm CCFA thin film deposited at 100$^{\rm o}$C on an Al$_2$O$_3$ (11$\bar{2}$0) substrate and annealed at 600$^{\rm o}$C. The inset shows the rocking curve of the (220) reflection.}
\end{figure}

\begin{figure}
\center
\includegraphics[width=300pt, angle=-90]{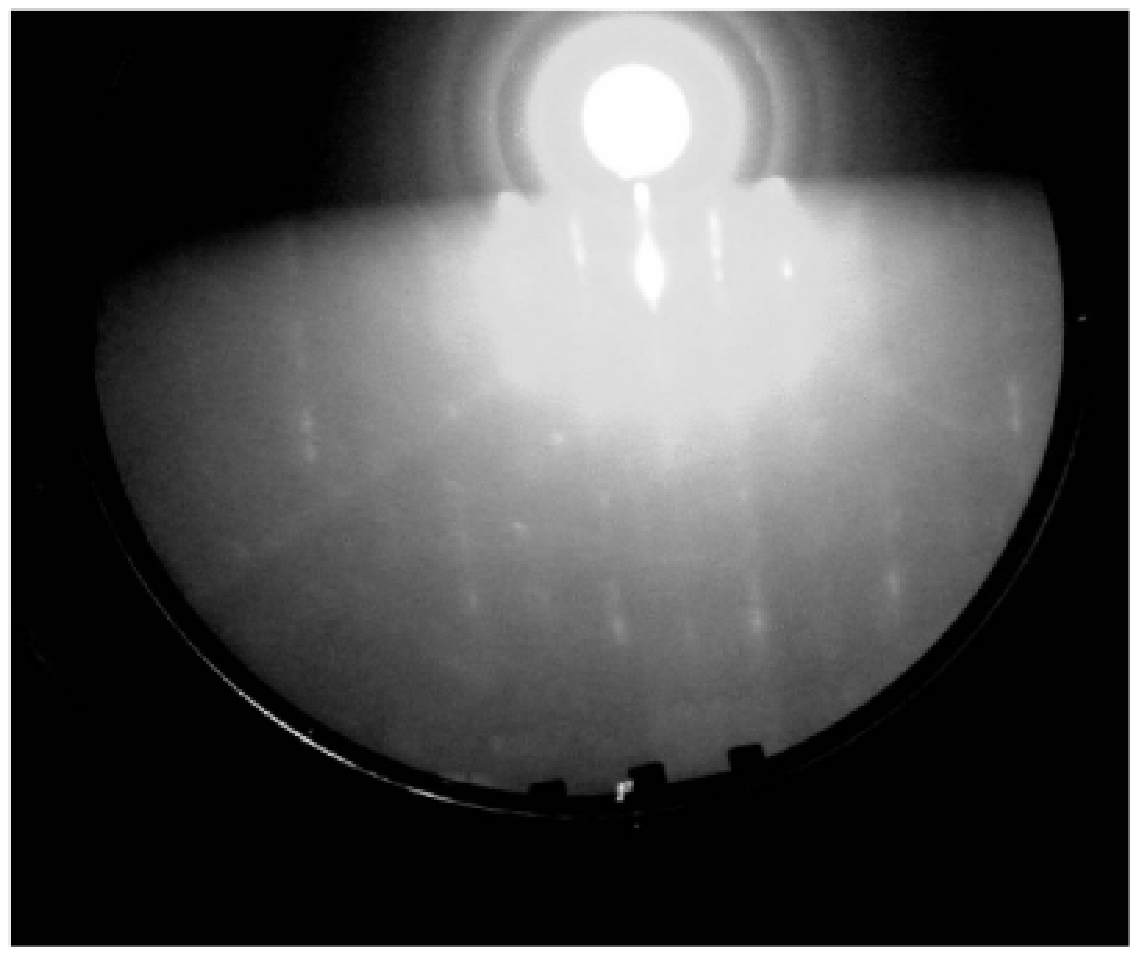}
\caption{\label{rheed-alo} RHEED pattern of a CCFA thin film deposited on  Al$_2$O$_3$ (11$\bar{2}$0) and annealed at 600$^{\rm o}$C.}
\end{figure}

\section{\label{sec:res}Structural properties}

The structural properties of the bulk of the CCFA thin films were studied by x-ray diffraction. The information about the degree of disorder of the samples is extracted by analysing the intensities of the (200) and (400) reflections. The off specular (111) reflection, which is characteristic for the fully ordered L2$_1$ structure of the Heusler compound, could not be observed (by 4-circle x-ray diffraction). 

In fig.\,\ref{mgo}, a x-ray $\Theta$/2$\Theta$ scan of CCFA thin film is shown. The film was deposited at 100$^{\rm o}$C on a Fe buffer layer (10 nm) on a MgO (100) substrate and then annealed at 600$^{\rm o}$C for 5 min. In fig.\,\ref{rheed-mgo} a diffraction pattern of high energy electrons at glancing incidence (RHEED) of such a CCFA thin film is shown, proving surface and in-plane order.
Please note that the CCFA phase can be observed already in the x-ray diffractogram of non-annealed samples, but the RHEED pattern is only visible with annealed samples. Additionally, compared to as deposited CCFA thin films \cite{conca}, the annealed samples show a clearly increased x-ray scattering intensity.

In fig.\,\ref{mgalo}, a x-ray scan of a CCFA thin film deposited directly on a different substrate, MgAl$_2$O$_4$ (100), is shown. In contrast to the films on a buffer layer on MgO substrates, now for the observation of any x-ray scattering intensity the annealing process of the thin film is required. However, the RHEED patterns of these annealed samples show Kikuchi lines (fig.\,\ref{rheed-mgalo}) which are indicating a reduced defect density at the film surface compared to the samples on buffer layers.
For a quantification of the disorder on the Co sites of the Heusler compound we consider the ratio of the intensities of the (200) and (400) x-ray reflections. For this analysis both intensities have to be multiplied by sin($\theta$) ($\theta$: angle of incidence of the x-ray beam on the sample surface) in order to correct geometrical thin film effects. This factor estimates the effect of the inclination of the sample with respect to the x-ray beam on the total intensity received by the thin film which is penetrated by most of the x-ray intensity without scattering. The corrected ratio amounts to a value larger than 0.27 for samples deposited on a buffer layer on MgO or directly on MgAl$_2$O$_4$. Comparing this value with a x-ray diffraction simulation (PowderCell) for different degrees of disorder, we conclude that in the annealed CCFA thin films the disorder on the Co sites is smaller than 10\%. In contrast, in non annealed samples (on MgO/buffer layer) the ratio has a value of $\sim$0.14, which is consistent with a disorder levels near 20\%. 

Epitaxial growth has also been  achieved for films deposited directly on Al$_2$O$_3$ (11$\bar{2}$0) substrates, see fig.\,\ref{alo} and  fig.\,\ref{rheed-alo}. In this case the films grow in a different orientation with the (110)-direction perpendicular to the substrate surface. Again Kikuchi lines are visible in the RHEED pattern. As with films on MgAl$_2$O$_4$, CCFA x-ray reflections are only observed after annealing of the sample. The width of the x-ray rocking curves ($\omega$-scans) is smaller for thin films deposited directly on MgAl$_2$O$_4$ and Al$_2$O$_3$ substrates than for films deposited on buffer layers on MgO substrates indicating a reduced mosaicity.

In contrast to other reports \cite{marukame}, a direct deposition on MgO substrates without any buffer layer did not result in epitaxial growth.

The width of the CCFA reflections in $\Theta$/2$\Theta$ scans gives additional information about the crystallographic quality of the films. Applying the Scherrer equation, the width of the reflections is related to the structural coherence length of the sample.
Using the substrate peaks to subtract the contribution of the finite angular resolution of the diffractometer, a coherence length of 50nm is extracted for a 100nm film. This value is an indication for long range structural coherence of the epitaxial films.

\section{\label{sec:mtj}Magnetic tunneling junctions}

This work is focused on MTJ's with amorphous AlO$_x$ barriers which were all prepared in the same way: The annealing of the CCFA electrode was done prior to the deposition of the Al layer which was oxidized for 120s or 90s in a mixture  of 50\% argon and 50\% oxygen to form the tunnelig barrier. Co was used as a well known ferromagnetic counter electrode. The Co film was oxidised softly in order to create an antiferromagnetic CoO top layer which is necessary for the exchange bias of the magnetization of the Co tunneling electrode. Finally, a thin layer (10 nm) of Pt was deposited on top of the junction to prevent surface degradation in air and to reduce the contact resistance. This stack of layers was patterned by photolitography to obtain  circular MESA structures with different diameters from 100$\mu$m up to 200$\mu$m.

\begin{figure}
\center
\includegraphics[width=300pt, angle=-90]{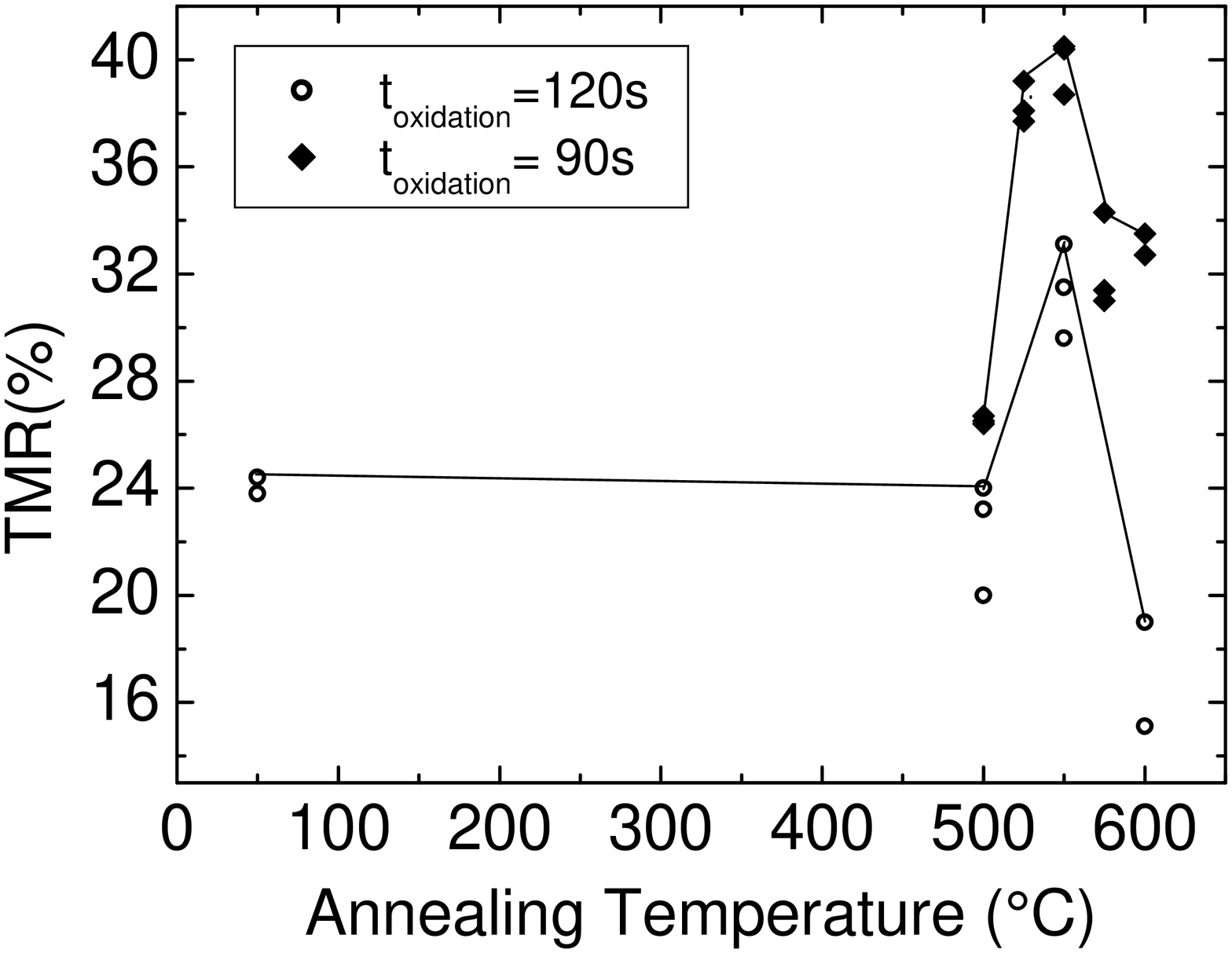}
\caption{\label{anneal} Effect of the annealing temperature of the CCFA electrode on the TMR ratio (annealing time 5 min). The points at 50$^{\rm o}$C represents non-annealed samples.}
\end{figure}

The effect of the annealing temperature on the TMR ratio of junctions with CCFA electrodes deposited on Fe buffer layers on MgO substrates was studied. As shown in fig.\,\ref{anneal}, the TMR shows an abrupt increase for annealing temperatures above 500$^{\rm o}$C.

\begin{figure}
\center
\includegraphics[width=300pt, angle=0]{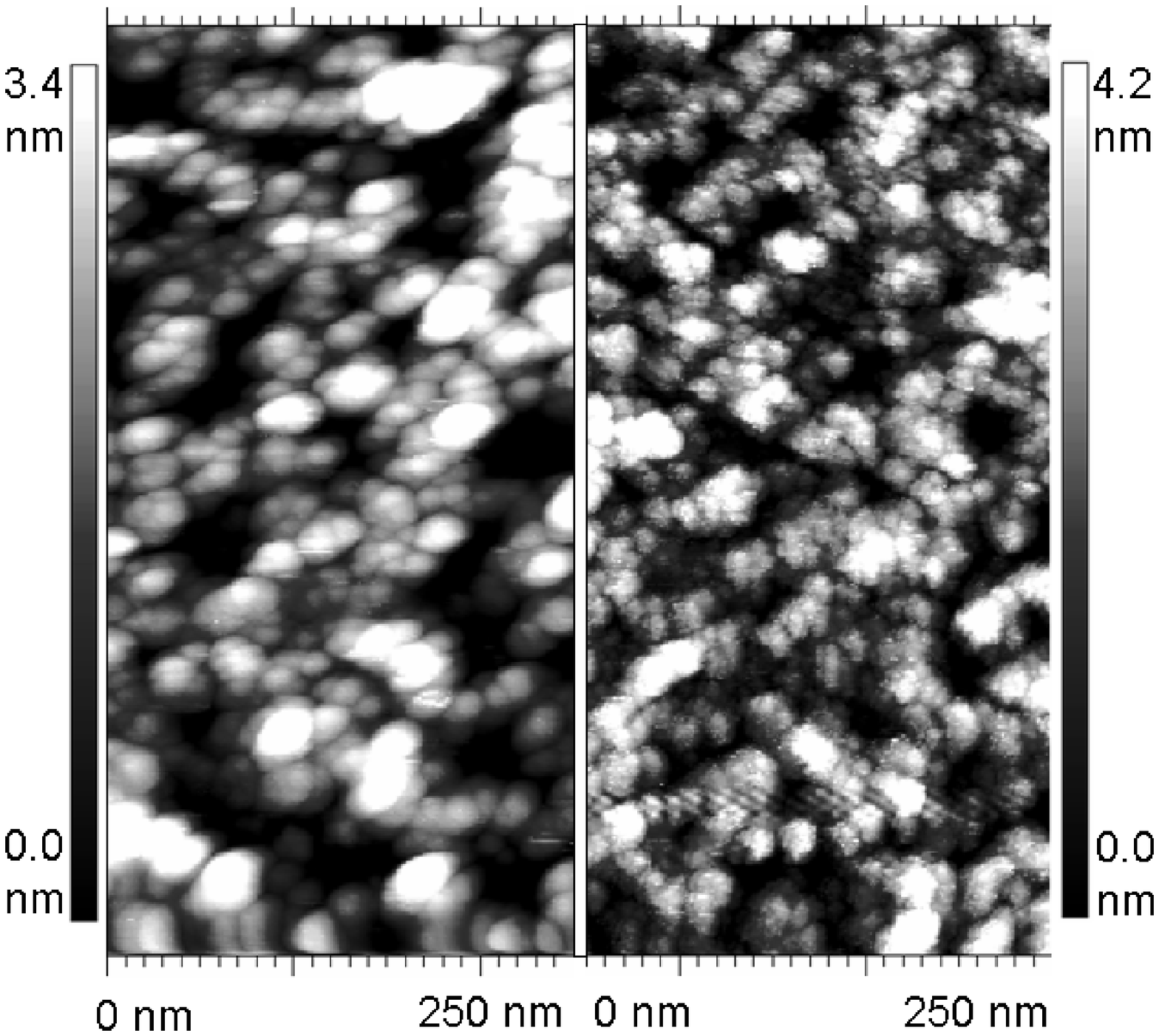}
\caption{\label{STM} In situ STM images of CCFA surfaces with different annealing temperatures (left: T$_{anneal}$ = 600$^{\rm o}$C, right: T$_{anneal}$ = 550$^{\rm o}$C}
\end{figure}

This increase coincides with the appearance of clear RHEED reflections arranged on circular arcs, which is characteristic for scattering at a 2-dimensional single crystalline surface (fig.\,\ref{rheed-mgo}). In contrast to these patterns the RHEED images observed at reduced annealing temperatures consist of more and irregularly arranged spots which indicates scattering at 3-dimensional structures. Thus the larger TMR is correlated with an improved surface order due to the increased annealing temperature.
However, further increasing the annealing temperature above 550$^{\rm o}$C results in a equally abrupt reduction of the TMR. Tunneling junctions with CCFA electrodes annealed at 575$^{\rm o}$C show already a clearly decreased TMR. The apparently obvious expectation that this reduction is due to rougher surfaces of the CCFA electrodes resulting in a reduced barrier quality could not be confirmed by in situ scanning tunneling microscopy (STM). In fig.\,\ref{STM} STM-images of CCFA samples annealed at 600$^{\rm o}$C and 550$^{\rm o}$C are shown. Both images show a similar morphology with a root mean square roughnesses of $1.1$nm (T$_{anneal}=600^{\rm o}$C) and $1.4$nm (T$_{anneal}=600^{\rm o}$C). Since XRD and RHEED indicate a further improved crystallographic quality of the CCFA with T$_{anneal}=600^{\rm o}$C, the reason for the TMR reduction remains to be clarified. Additional investigations, e.\,g.\,of the chemical composition of the CCFA at the surface, are necessary.
   
\begin{figure}
\center
\includegraphics[width=300pt, angle=-90]{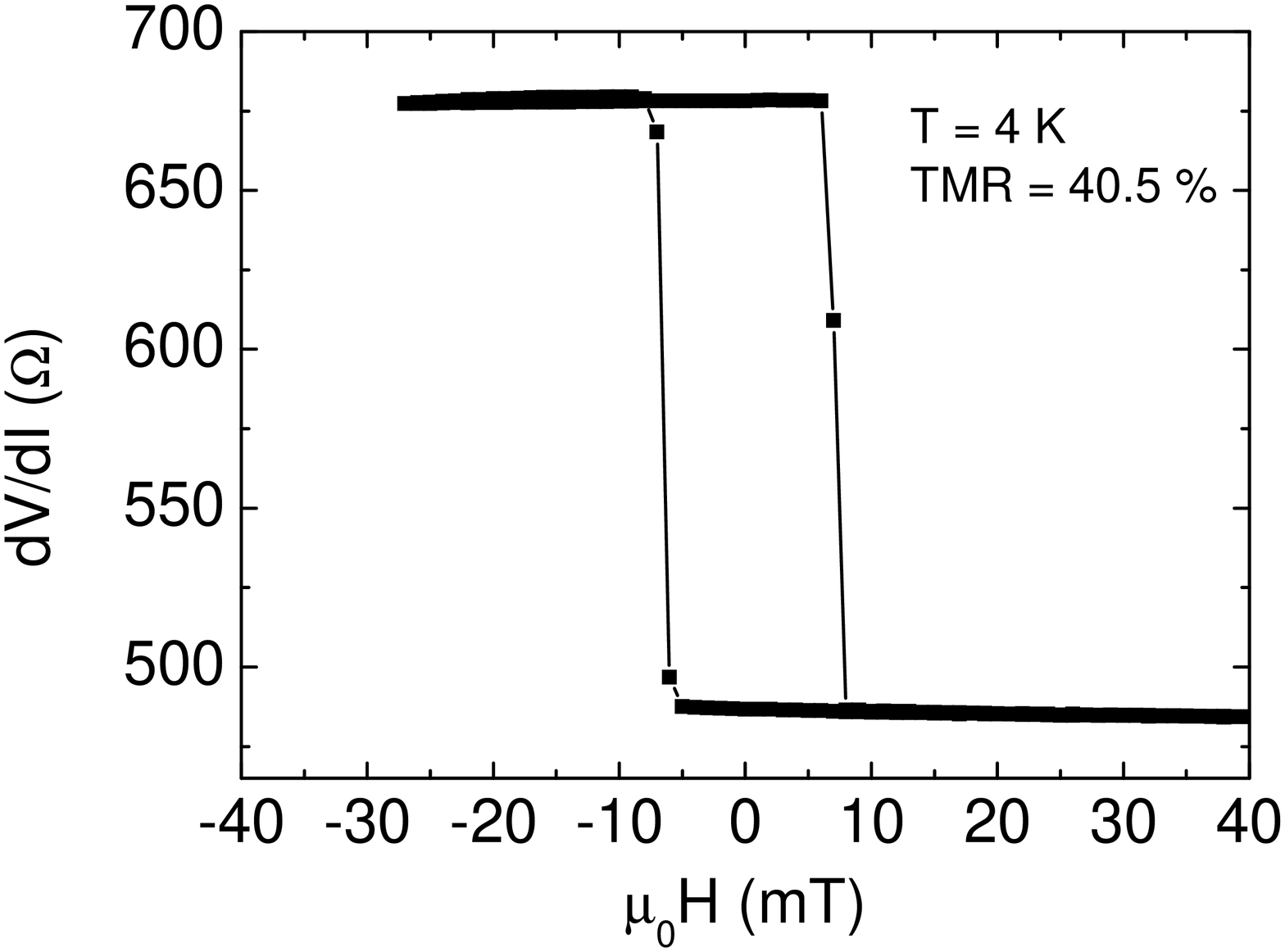}
\caption{\label{tmr} Magnetoresistance of a MgO$\backslash$Fe$\backslash$CCFA$\backslash$AlO$_x$$\backslash$Co\\$\backslash$CoO$_x$$\backslash$Pt tunneling junction (V$_{bias}$=1mV).} 
\end{figure}

The highest observed TMR ratio amounts to $40.5\%$ at 4K and $34.6\%$ at 77K (fig.\,\ref{tmr}). From Co$\backslash$AlO$_x$$\backslash$Co$\backslash$CoO reference tunneling junctions a spin polarisation of 31\% was deduced for our Co electrodes. Assuming this value we obtain from the Julli\`ere model a spin polarisation of 54\% for CCFA. This result was achieved with a CCFA electrode deposited on a Fe buffer layer and annealed at 550$^{\rm o}$C.\\
Typically the TMR values obtained from different junctions prepared at the same conditions scatter in a range of $\pm 2\%$ around a mean value. Sporadic strongly reduced TMR values can be associated with problems in the pattering process.

\begin{figure}
\center
\includegraphics[width=300pt, angle=-90]{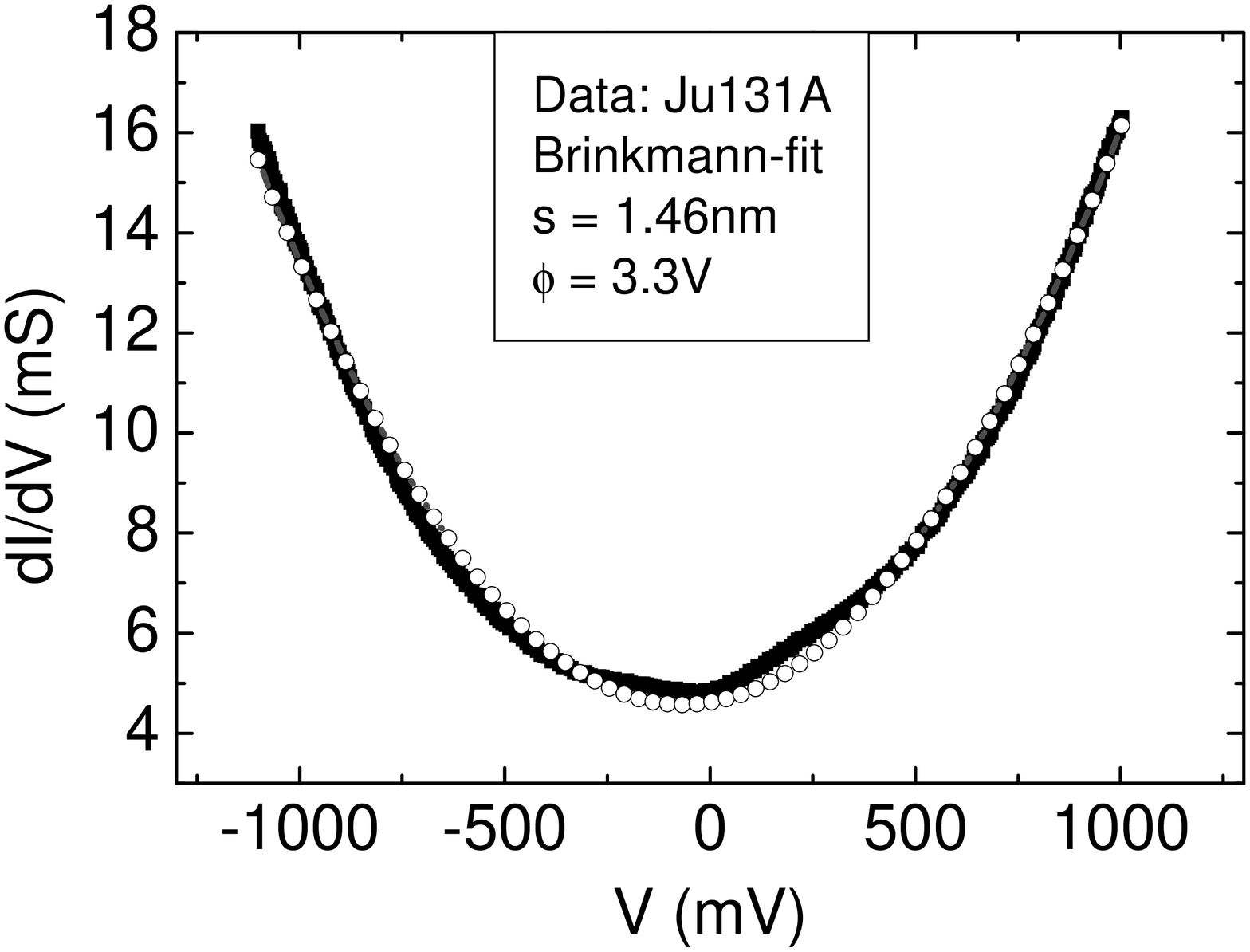}
\caption{\label{brinkm} Bias voltage dependence of the differential conductivity of a MgO$\backslash$Fe$\backslash$CCFA$\backslash$AlO$_x$$\backslash$Co$\backslash$Pt tunneling junction. Solid points are the measured data. The empty circles represent a fit using the Brinkmann model. The averaged height ($\phi$) and effective width ($s$) are given in volts and \aa ngstr\"om, respectively.}
\end{figure}

Typical samples show a linear increase of the resistance of about 30\% with decreasing temperature down to 4K. We consider this linear behaviour to be a fingerprint of a well defined tunneling barrier. According to our knowledge the highest published spin polarisation of Co$_2$Cr$_{0.6}$Fe$_{0.4}$Al as deduced from the TMR of tunneling junctions with AlO$_x$ barrier and applying the Julli\`ere model was obtained by Inomata et al.\,\cite{inomata}. These authors calculated a spin polarisation at 5K of 52\%, which is close to the 54\% we deduced using a different counter electrode in the MTJ's.
For the characterization of the tunneling barriers we measured the bias voltage dependence of the differential conductivity of the MTJ's for parallel electrode magnetization (fig.\,\ref{brinkm}). From the Brinkmann model \cite{brinkmann}, which is based on the assumption of a square potential of the barrier which is deformed trapezoidally by the bias voltage, the average height $s$ and effective width $\phi$ of the barrier potential are extracted. The value of $s\simeq 1.5$nm tends to be smaller than the AlO$_x$ barrier thickness expected from our deposition rate. Thus an increase of the TMR effect by a reduction of the thickness of the Al layer of our junctions seems to be feasible.

Although the CCFA thin films deposited directly on MgAl$_2$O$_4$ and Al$_2$O$_3$ showed improved RHEED reflections compared to the samples on MgO with buffer layers, the observed TMR ratios of junctions with those base electrodes were reduced. For samples annealed at 550$^{\rm o}$C, maximum TMR ratios of 24\% on MgAl$_2$O$_4$ and 19\% on Al$_2$O$_3$ were observed.

\section{\label{sec:con}Conclusions}

Epitaxial Co$_2$Cr$_{0.6}$Fe$_{0.4}$Al thin films can be grown in the B2 structure with only a small partition of disorder on the Co sites. This is possible due to an appropriate annealing procedure on different substrates and buffer layers. Additionally the annealing procedure generates a well ordered thin film surface. Appropriate annealing of the Co$_2$Cr$_{0.6}$Fe$_{0.4}$Al thin film improves the structural quality of this electrode and increases the TMR effect. Using a Co counter electrode for the tunneling junctions we deduce from the Julli\`ere model a spin polarisation of similar magnitude $\simeq 54\%$ as obtained before by Inomata et al.\,\cite{inomata} employing a Co$_{75}$Fe$_{25}$ counter electrode. With this result the spin polarisation of the Heusler compound is still far from the theoretically expected 100\%. This may be associated with interface and surface effects which influence the band structure \cite{galanakis}, the limited validity of the Julli\`ere model for tunneling \cite{maclaren}, or the more technical optimization of the tunneling barrier. On the other hand, with a spin polarisation of 54\% as deduced by tunneling with AlO$_x$ barriers, the highest spin polarisation values of conventional ferromagnets \cite{monsma} are already approached.\\
{\bf acknowledgments}\\
This project is financially supported by the {\em Stiftung Rheinland-Pfalz f\"ur Innovation}, project no. 699.


\section*{References}
\begin{thebibliography}{10}
\bibitem{fecher} Fecher G, Kandpal H, Wurmehl S, Morais J, Lin H J, Elmers H J, Sch\"onhense G and Felser C 2002 {\em J.\,Phys: Condens.\,Matter} {\bf 17} 7237
\bibitem{wurmehl} Wurmehl S, Fecher G, Kroth K, Kronast F, D\"urr H, Takeda Y, Saito Y, Kobayashi K, Lin H J, Sch\"onhese G, and Felser C 2006 {\em J.\,Phys.\,D: Appl.\,Phys.} {\bf 39} 803
\bibitem{julliere} Julli\`ere M 1975 {\em Phys.\,Lett.\,A} {\bf 54} 225
\bibitem{conca} Conca A, Jourdan M, Herbort C and Adrian H 2006 {\em preprint:\,cond-mat/0605698}, {\em J.\,Crystal Growth}, accepted
\bibitem{marukame} Marukame T, Ishikawa T, Matsuda K, Uemura T and Yamamoto M 2006 {\em Appl.\,Phys.\,Lett.} {\bf 88} 262503-1
\bibitem{mavropoulos} Mavropoulos P, Papanikolaou N and Dederichs P H 2000 {\em Phys.\,Rev.\,Lett.} {\bf 85} 1088
\bibitem{butler} Butler W H, Zhang X G, Schulthess T C and MacLaren J M 2001 {\em Phys.\,Rev.\,B} {\bf 63} 054416
\bibitem{inomata} Inomata K, Okamura S, Miyazaki A, Kikuchi M, Tezuka N, Wojcik M and Jedryka E 2006 {\em J.\,Phys.\,D: Appl.\,Phys.} {\bf 39} 816
\bibitem{miura} Miura Y, Nagao K and Shirai M 2004 {\em Phys.\,Rev.\,B} {\bf 69} 144413
\bibitem{brinkmann} Brinkmann W F, Dynes R C and Rowell J M 1970 {\em J.\,Appl.\,Phys.} {\bf 41} 1915
\bibitem{galanakis} Galanakis I 2002 {\em J.\,Phys: Condens.\,Matter} {\bf 14} 6329
\bibitem{maclaren} MacLaren J M, Zang X G, Butler W H 1997 {\em Phys.\,Rev.\,B} {\bf 56} 
\bibitem{monsma} Monsma D J and Parkin S S P 2000 {\em Appl.\,Phys.\,Lett.} {\bf 77} 720 11827
\end{thebibliography}
\end{document}